\documentstyle[12pt]{article}
\newcommand{\be}{\begin{equation}}
\newcommand{\ee}{\end{equation}}
\newcommand{\ben}{\begin{eqnarray}}
\newcommand{\een}{\end{eqnarray}}
\begin{document}

\begin{center}
{\bf Exact Kink Solitons in the Presence of Diffusion,\\
        Dispersion, and Polynomial Nonlinearity\footnote{This work
is supported in part by funds provided by the U. S. Department of Energy
(D.O.E.) under cooperative research agreement DE-FC02-94ER40818, and by
Conselho Nacional de Desenvolvimento Cient\'\i fico e Tecnol\'ogico,
CNPq, Brazil.}}
\end{center}

\begin{center}
E.P. Raposo$^{a,}$\footnote{Corresponding author. Present address: 
Departamento de F\'\i sica, Universidade
Federal de Pernambuco, 50670-901 Recife, Pernambuco, Brazil. 
Telephone: 55 81 271.0111. Fax: 55 81 271.0359. 
E-mail: raposo@cmt.harvard.edu or ernesto@lftc.ufpe.br ~.} and 
D. Bazeia$^{b,}$\footnote{On leave from Departamento de F\'\i sica, 
Universidade
Federal da Para\'\i ba, Caixa Postal 5008, 58051-970 Jo\~ao Pessoa,
Para\'\i ba, Brazil.} 
\end{center}

\begin{center}
$^{a}$Lyman Laboratory of Physics\\
Harvard University, 
Cambridge, Massachusetts 02138\\

\vskip .5cm

$^{b}$Center for Theoretical Physics\\
Laboratory for Nuclear Science and Department of Physics\\
Massachusetts Institute of Technology, Cambridge, Massachusetts 02139-4307
\end{center}

\vskip 1.0cm

\begin{center}
(MIT-CTP-2742, May 1998)
\end{center}

\vskip 1.5cm

\begin{center}
Abstract
\end{center}
We describe exact kink soliton solutions
to nonlinear partial differential equations in the generic form 
$u_{t} + P(u) u_{x} + \nu u_{xx} + 
\delta u_{xxx} = A(u) $, with polynomial functions $P(u)$ and $A(u)$ of 
$u=u(x,t)$, whose generality allows the identification 
with a number of relevant equations in physics. 
We emphasize the study of chirality of the solutions, 
and its relation with diffusion, dispersion, 
and nonlinear effects, as well as its dependence on the 
parity of the polynomials $P(u)$ and $A(u)$ with respect 
to the discrete symmetry $u\to-u$. We analyze two types 
of kink soliton solutions, which are also solutions
to $1+1$ dimensional $\phi^{4}$ and $\phi^{6}$ field theories. 

\vskip .8cm

\begin{center}
\end{center}

\begin{center}
\end{center}

\newpage

\renewcommand{\thesection}	{\Roman{section}}

\section{Introduction}

Nonlinear differential equations are known \cite{whi74} to describe 
a wide variety of phenomena not only in physics, in which applications 
extend over magnetofluid dynamics, water surface 
gravity waves, electromagnetic radiation reactions, and ion acoustic waves
in plasmas, just to name a few, 
but also in biology and chemistry, among several other fields. In spite of the
increasing development of mathematical techniques and concepts to solve 
nonlinear equations, 
exact solutions seem not to be the rule and numerical methods 
have been the most common approach to study their properties \cite{num}.

Since the seminal work of Korteweg and de Vries (KdV) \cite{kdv},
in which a third-order 
nonlinear equation was studied to explain the shallow-water 
solitary wave experiments by Russel \cite{rus}, 
solitary-wave or soliton solutions have been found in a number of 
nonlinear differential equations \cite{whi74}, 
with phenomenologies related to liquid crystals, 
dynamics of growing interfaces and domain walls, 
eletromagnetism, nonlinear optics, acoustics, and elasticity, among others. 
These localized non-dispersive travelling waves can be of several 
distinct types \cite{kink}, such as kink, pulse, breather, envelope, and dark 
solitons, all of them presenting the property that 
their shapes and velocities are preserved along 
propagation, and even upon collision with 
other solitary waves. In fact, they result from a precise balance 
among the competing elements that define the equation, namely the tendency 
of spreading due to the presence of a dispersive term and the action 
of nonlinear terms, which in general favour large amplitude disturbances 
and velocities that assure the stability of the travelling waves 
with respect to small 
distortions in form. In some special cases, solitary waves do not 
support arbitrary speeds and can propagate only with definite 
velocities determined from the parameters of the nonlinear equation. 
These so-called chiral solitons have received 
much attention lately due to their recent associations with the nonlinear 
Schr\"odinger equation and the fractional quantum Hall effect
\cite{ben96,jac96,gse97}. Moreover, we have also recently pointed \cite{ern} 
to the presence of chiral kink soliton solutions in generalized 
KdV-Burgers-Huxley (gKdVBH) equations in the form
\be
u_t+f_x+g_{xx}-\bar{\delta}\,u_{xxx}= h(u) ~,
\ee
where $f$, $g$, and $h$ are smooth functions in $u$, and 
$u_{t}$ ($u_{x}$) stands for the partial derivative of $u(x,t)$ 
with respect to time (position). 
The above equation combines dispersion, 
controlled by the real parameter $\bar{\delta}$, nonlinearity, described
by the functions $f(u)$, $g(u)$, and $h(u)$, and diffusion
related to the second-order space derivative.

This work is inspired in our previous study \cite{ern}, although here we
follow a different approach. We consider exact kink solutions to nonlinear
differential equations of the generic form
\be
u_{t} + P(u)\, u_{x} + \nu\, u_{xx} + 
\delta\, u_{xxx} = A(u)~,
\ee
with polynomial functions defined as
\be
P(u) = \sum_{i=0}^{N_{p}} p_{i}\, u^{i}~,
\ee
and
\be
A(u) = \sum_{i=0}^{N_{a}} a_{i}\, u^{i}~.
\ee
The general form of Eq.~(2) allows the identification of several
interesting cases  \cite{whi74}.
For instance, by considering propagating waves 
$u(x,t)=u(x-ct)=u(y)$, with velocity $c$, it is easy to see that the
gKdVBH equation is recovered from Eq.~(2) for $df/du =P(u)$,
$g(u)=g_{0}+\nu u$, $\bar{\delta}=-\delta$, and $h(u)=A(u)$. Furthermore,
the standard KdVB equation corresponds to identifying $P(u)=\lambda u$,
$\delta=-\bar{\delta}$, and $A(u)= 0$, and the modified KdVB equation requires
$P(u)=3\bar{\lambda} u^{2}$, $\delta= -\bar{\delta}$, and $A(u)=0$,
with the particular case $\nu = 0$ accounting respectively for the standard
and modified KdV equations. On the other hand, the Burgers-Huxley equation
represents the situation in which $P(u)=\bar{\lambda} u$,
$\delta=0$, and $A(u)=h(u)$, with the case $h(u)=0$ corresponding to the 
standard Burgers equation, whose applications include nonlinear 
heat diffusion, shock waves, viscous effects in gas dynamics, and an
important connection with the deterministic KPZ equation \cite{kpz}
in one spatial dimension, known to provide the evolution of the profile
of a growing interface or a domain wall of general nature.

Other examples include the
generalized Boussinesq equation \cite{whi74,kink},
\be
u_{tt} - c_{0}^{2} u_{xx} - p (u^{2})_{xx} - q (u^{3})_{xx} - 
\bar{h} u_{xxxx} = 0~,
\ee
which by a trivial integration in $y$ for travelling solutions 
$u(x,t)=u(x-ct)=u(y)$
is related to Eq.~(2) through the identifications
$P(u) = -c_{0}^{2} - 2 p u - 3 q u^{2}$, $\nu = 0$, $\delta = - \bar{h}$, 
$A(u) = a_{0}$, where $c_{0}$ is the speed of the sound. The
standard and modified
Boussinesq equations correspond respectively to
the cases $q = 0$ and $p = 0$. They were
first described in the context of the
theory of long water waves, but also present
innumerous applications in electromagnetism, plasma physics, elasticity,
and other fields. Another generic equation commonly found in physics 
is \cite{whi74}
\be
u_{tt} - \alpha^{2} u_{xx} + \frac{dV}{du} = 0~,
\ee
which can also be compared to Eq.~(2) by considering travelling waves with
$y = x - \bar{c}\,t$ and identifying $P(u) = c$,
$\nu = \bar{c}^{2} - \alpha^{2}$, $\delta = 0$, 
and $A(u) = - (dV/du)$. As examples of Eq.~(6), we recall the Klein-Gordon 
equation of field theory in which $V(u) = \beta^{2} u^2$, also 
applied to describe standard vibrations for a 
displacement $u$ with presence of an additional restoring force 
proportional to $u$, and the sine-Gordon equation corresponding to 
$V(u) = \beta^{2} (1-\cos u)$, which embodies a wide range of applications
including Josephson junctions in superconductors, dislocations in crystals,
waves in ferromagnetic materials, laser pulses in two state media,
and geometry of surfaces, just to name a few.
Also, the standard nonlinear Schr\"odinger equation,
\be
iu_t+\mu\,u_{xx}=\frac{dV}{d\rho}\,u~,
\ee
corresponds to formally assigning for the travelling wave $u(x-\bar{c}\,t)$
the values $P(u) = c - i\bar{c}$, $\nu = \mu$, $\delta = 0$,
and $A(u) = (dV/d\rho)\, u$, where $V=V(u^*u)=V(\rho)$ is commonly
expressed as a quadratic or cubic potential in the charge density $\rho$.

Regarding the polynomial $A(u)$ in Eq.~(2),
it first appeared as a generalization \cite{kink}
of the study of monoatomic chains of
equal masses interacting via a number of nearest-neighbor realistic 
potentials such as 
the Toda, Morse, and Lennard-Jonnes potentials \cite{toda}.
In the continuum spatial
limit, the longitudinal displacement of the particle $n$ from its
equilibrium position, $u_{n}(t)$, is replaced by an analytic function
$u(x,t)$, and the series expansion of the
interaction terms gives rise to the spatial derivatives
of high orders.
In this context, $A(u)$ thus represents an on-site local interaction
potential in the differential equation of motion that governs the
system \cite{kink}. 

In the following we generalize the procedure applied in Ref.~\cite{han}
to some nonlinear equations, in order to investigate Eq.~(2).
The extension to equations with higher-order derivative terms, such as
the fifth-order KdV-type equations \cite{num}, is straightforward and will not 
be considered here.
Our emphasis will be on
the study of chirality of exact kink solutions to the general equation,
and analysis of its relation with combined diffusion, dispersion,
and nonlinear effects, as well as its dependence on the
parity of the polynomials $P(u)$ and $A(u)$ with respect
to the discrete symmetry $u\to-u$. In Sec.~II we analyze the case
in which kink solutions are of the type $u(y) = a \tanh(\lambda a y)$,
whereas in Sec.~III solutions in the form
$u(y) = \{ (a^{2}/2) [ 1 + \tanh (\lambda a^{2} y ) ] \}^{1/2}$
are considered. Finally, comments and conclusions are presented in Sec.~IV.

\section{Exact Kink Solutions of Type $\phi^4$}

We start by considering travelling solutions $u(x,t)=u(x - ct)=u(y)$ to 
the general third-order nonlinear differential equation in the
presence of a local polynomial potential $A(u)$, Eq.~(2), 
\be
[-c +P(u)] \frac{du}{dy}+\nu \frac{d^{2}u}{dy^{2}}+ 
\delta \frac{d^{3}u}{dy^{3}}=A(u)~.
\ee
From Eqs.~(3) and (4), the orders of the polynomials, $N_{p}$ and $N_{a}$,
are not arbitrary, but are instead strongly related to 
the characteristics of the
solutions. For instance, if
$N_{p} = N_{a} - 2$ one can search for exact
kink soliton solutions in the form,
\be
u(y) = a \tanh(\lambda a y)~,
\ee 
since the property,
\be
\frac{du}{dy} = \lambda (a^2 - u^2)~,
\ee
leads to the general result,
\be
{\cal O}( \frac{d^{n}u}{dy^{n}} ) = u^{n+1}~.
\ee
Indeed, from Eq.~(11) one sees that if $N_{a} > N_{p} + 2$ no
solutions of the form presented in Eq.~(9) can be found to Eq.~(8).
We have shown \cite{ern,baz} that Eq.~(9) also corresponds to kink
solutions $\phi = \phi(y)$ in the context of the relativistic $\phi^{4}$ field
theory in $1+1$ dimensions, which presents potential
\be
V(\phi)=\frac{1}{2}\lambda^2(\phi^2-a^2)^2~,
\ee
that develops spontaneous symmetry breaking of the discrete $Z_2$ symmetry
$\phi\to-\phi$. Here we mention that the kink solutions resembles 
the two asymmetric and degenerate vacua of this $\phi^4$ system.

The coefficients $\{p_{i}\}$ and $\{a_{i}\}$ in Eqs.~(3) and (4) are not
arbitrary, but instead must obey a series of relations in order to allow
Eq.~(9) to be solution of Eq.~(8). In fact, this is actually expected
to occur to solitary-wave solutions, which can only exist for a precise
balance of the diffusion, dispersion, and nonlinear effects. 
One can see that by explicitly substituting Eqs.~(3), (4), and
(9) in Eq.~(8), to obtain
\ben
\frac{a_{0}}{\lambda a^2}&=&-c + p_{0}-2 \delta \lambda^{2} a^{2}~,\\
\frac{a_{1}}{\lambda a^2}&=&p_{1} - 2 \nu \lambda~,\\
\frac{a_{2}}{\lambda a^2}&=&p_{2} + 8 \delta \lambda^{2} +
\frac{(c - p_{0})}{a^{2}}~,\\
\frac{a_{3}}{\lambda}&=&-p_{1} + 2 \nu \lambda~,\\
\frac{a_{4}}{\lambda}&=& -p_{2} - 6 \delta \lambda^{2}~.
\een
Although this technique can be easily generalized
to any polynomial orders $N_{p}$ and $N_{a}$, provided that in the case of
solution given by Eq.~(9) they obey the relation $N_{p} = N_{a} - 2$,
we have restricted ourselves in Eqs.~(13)-(17) to situations of
physical interest, representing quadratic nonlinearity in the
first-order derivative term, $N_{p} = 2$, as pointed from the examples
listed in Sec.~I.
Indeed, considering the general $N_{p} > 2$, $N_{a} > 4$ case only generates
relations among highest-order coefficients $\{p_{i}\}$ and $\{a_{i}\}$
not involving the parameters of interest 
$\nu$, $\delta$, and $c$. For instance,
in this case Eqs.~(16) and (17)
would respectively read
\ben
\frac{a_{3}}{\lambda}&=&-p_{1} + a^2 p_{3} + 2 \nu \lambda~,\\
\frac{a_{4}}{\lambda}&=&-p_{2} + a^2 p_{4} - 6 \delta \lambda^{2}~,
\een
and new relations involving higher-order coefficients would emerge:
\be
\frac{a_{i}}{\lambda} =  -p_{i-2} + a^2 p_{i} \;\;,\;\; n > 4~.
\ee
Notice that Eq.~(20) represents matches
exclusively between coefficients of the polynomials $P(u)$ and $A(u)$
necessary in order to allow Eq.~(9) as solutions to Eq.~(8). No extra
physical information
concerning diffusion or dispersion effects, or even chirality (see below)
is gained from Eq.~(20), contrarily to what happens
to Eqs.~(13)-(17), as we discuss below. Moreover, Eqs.~(13)-(20)
also indicate that the coefficients are not totally independent,
but instead must obey a set of relations, which in the particular case of
solutions given by Eq.~(9) are given by, in the case of $N_{a}$ even,
\be
\sum_{i = 0}^{N_{a}/2 - 1} a^{2i} a_{2i + 1} = 0~,
\ee
and
\be
\sum_{i = 0}^{N_{a}/2} a^{2i} a_{2i} = 0~.
\ee
Consequently, setting the values of the $N_{a}/2$ ($N_{a}/2 - 1$) even 
(odd) $a_{i}$'s automatically determines the values of the remaining
even (odd) coefficients. 

We notice from Eqs.~(13)-(20) 
that the diffusion and dispersion parameters, respectively
$\nu$ and $\delta$, are related to definite and distict parities of the 
coefficients. For instance, $\nu$ is associated only with odd coefficients,
$p_{1}$, $p_{3}$, $a_{1}$, $a_{3}$, whereas $\delta$ is related to the
even ones, $p_{0}$, $p_{2}$, $p_{4}$, $a_{0}$, $a_{2}$, $a_{4}$.
No higher-order coefficients depend on $\nu$ or $\delta$.
Indeed, this is consequence of properties of the solution 
considered, Eqs.~(9) and (10), 
namely that each derivative multiplies the original 
function by a factor of order $u$, thus changing its parity with respect
to the symmetry $u\to-u$, as one can also see from Eq.~(11).
Another important feature is that the velocity $c$ of
the kink solitons also depends only on even coefficients up to quadratic 
order, i.e., $p_{0}$, $p_{2}$, $a_{0}$, $a_{2}$. The fact that $c$ 
is not arbitrary, being instead determined by the parameters that 
define the nonlinear differential equation, indicates that solutions given 
in terms of Eq.~(9) are actually chiral solitons. For instance, Eq.~(13) can
be used to determine the velocity of the localized travelling solutions,
\be
c = p_{0}-\frac{1}{\lambda a^{2}} ( a_{0}+ 2\delta \lambda^{3} a^{4} )~.
\ee
We observe from the above equation 
that $p_{0}$ plays the role of the velocity of the 
reference frame, as expected since it represents the $u$-independent 
term in the polynomial $P(u)$. 
Furthermore, Eq.~(23) also tells that 
the diffusion term associated with the parameter $\nu$ does not 
play any role in fixing the velocity of the solitons, thus not 
contributing to their chirality. 
This is a consequence of the distinct parities of the coefficients 
associated with 
$c$ and $\nu$, as commented above. Indeed, by considering the 
simplest case in which $p_{0} = a_{0} = 0$, we 
point to the relevance of 
the presence of a dispersive media ($\delta \ne 0$) in order to support 
chiral kink solitons of the type described by Eq.~(9). Nevertheless, 
this does not seem to be a property shared by all types of kink 
solitons, as we comment on Sec.~III. 

From the above discussions we also notice the remarkable fact that distinct
choices of the parameters can lead
to solitary-wave solutions with the same shape, Eq.~(9), even if they
do not necessarily have the same velocity. Indeed, this feature was
first verified \cite{toda} with the finding that Toda, Morse, and
Lennard-Jones solitons arising from distinct nonlinear differential
equations are very nearly alike, a result then attributed to the
similarity in the shapes of their respective potential walls.

Let us now consider some specific examples. The connection with the
gKdVBH equation, Eq.~(1) with the {\it ansatz} $u(x,t) = u(x - ct) = u(y)$,
is easily established from the following 
identifications (see also Introduction):
\ben
P(u)&=&\frac{d f(u)}{du}~,\\
\nu u &=& g(u) - g_{0}~,\\
\delta &=& - \bar{\delta}~,\\
A(u) &=& h(u)~,
\een
where $g_{0}$ is some unimportant constant.
To illustrate with a previously reported \cite{ern}
example of soliton solutions related to Eq.~(1) and with
form given by Eq.~(9), let us consider the case in which
\ben
P(u)& =& p_{0} + p_{2} u^{2}~,\\
\nu &=& 0 ~,\\
A(u) &=& a_{0} + a_{2} u^{2}~,
\een
that corresponds to the gKdVH equation with functions
\ben
f(u)&=&f_{0} +p_{o} u +\frac{ p_{2} u^{3} }{3}~,\\
g(u) &=& - g_{0}~,\\
h(u)& =&a_{0} + a_{2} u^{2}~.
\een
Solving Eqs.~(13)-(17) along with the functions expressed by Eqs.~(28)-(30)
leads to chiral solitons, Eq.~(9), with velocity
\be
c = p_{0} - \frac{a_{o} p_{2}}{ 3 a_{2} } +
\mbox{sgn}(\lambda) a_{2} \sqrt{ \frac{ -6 \delta }{p_{2}} }~.
\ee
The result reported in Ref.~\cite{ern} is restored by 
considering $p_{0} = - 2 \alpha^{2} \beta^{2}$, 
$p_{2} = 6 \beta^{2}$, $a_{0} = \alpha^{2} \beta$, and $a_{2} = -\beta$. 
In this case the velocity of the chiral solitons are simply given by 
($\delta < 0$)
\be
c = \mbox{sgn}(\lambda) \sqrt{-\delta\,}~.
\ee

\section{Exact Kink Solutions of Type $\phi^6$}

We now search for exact kink soliton solutions to
Eq.~(2) in the form
\be
u(y)=\{(a^2/2) [ 1+\tanh (\lambda a^{2} y ) ] \}^{1/2}~.
\ee
As we have reported in Refs.~\cite{ern,baz}, differently from the 
solutions investigated in Sec.~II, 
Eq.~(35) corresponds to kink solitons of the $\phi^{6}$
field theoretical system in the $1+1$ dimensional spacetime.
In this case the potential is
\be
V(\phi)=\frac{1}{2}\lambda^2\phi^2(\phi^2-a^2)^2~,
\ee
which also 
develops spontaneous symmetry breaking of the discrete $Z_2$ symmetry
$\phi\to-\phi$. Here, however, the kink solutions resembles both the
symmetric and asymmetric phases this $\phi^6$ system engenders.

In the present case we get
\be
\frac{du}{dy} = \lambda u (a^2 - u^2)~,
\ee
so that, by contrasting with the solutions studied in Sec.~II, 
\be
{\cal O}( \frac{d^{n}u}{dy^{n}} ) = u^{2n+1}~,
\ee
thus implying that each derivative operated preserves parity with 
respect to $u\to-u$ by multiplying the original function by an even factor on
$u$. As a consequence, one must now have $N_{p} = N_{a} - 3$ 
to allow Eq.~(36) to be solutions to Eq.~(8). In this case, however, 
since the third-order derivative of Eq.~(36) gives rise to a polynomial of 
order $u^{7}$, then the matching procedure involving physical
ingredients such as diffusion and/or dispersion requires
the presence of higher-order nonlinearity in $A(u)$, $N_{a} = 7$.
By substituting Eq.~(36) in Eq.~(8), we obtain the following
relations
\ben
a_{0}&=&0~,\\
\frac{a_{1}}{\lambda a^{2}}&=&-c+p_{0}+\lambda a^{2}\nu+
\lambda^{2} a^{4} \delta~,\\
a_{2}&=&\lambda a^{2} p_{1}~,\\
\frac{a_{3}}{\lambda}&=&c-p_{0}+a^{2} p_{2}-4 \lambda a^{2} \nu - 
13 \lambda^{2} a^{4} \delta~,\\
\frac{a_{4}}{\lambda}&=&-p_{1}+a^{2} p_{3}~,\\
\frac{a_{5}}{\lambda}&=&-p_{2}+a^{2} p_{4}+3 \lambda \nu +
27 \lambda^{2} a^{2} \delta~,\\
a_{6}&=&-\lambda p_{3}~,\\
\frac{a_{7}}{\lambda}&=&-p_{4}-15\lambda^{2}\delta~.
\een
First we notice that due to Eq.~(39) all terms in the left-hand side
of Eq.~(8) depend on $u$, thus leading to the result of Eq.~(40). 
Furthermore, as another consequence of Eq.~(39),
the diffusion and 
dispersion parameters, respectively $\nu$ and $\delta$, are 
{\it both} related only to even 
coefficients $\{p_{i}\}$ and 
odd coefficients $\{a_{i}\}$, differently from the case
considered in Sec.~II in which they were associated with coefficients 
of definite but distinct parities 
(even for $\delta$, odd for $\nu$). Consequently, the
physics related to diffusion and dispersion effects in Eq.~(8)
with solutions provided by Eq.~(36) is irrespective to the presence of odd
parity terms in the polynomial $P(u)$, and even terms in 
$A(u)$. Indeed, in this case the nonlinear differential
equation, Eq.~(8), presents odd symmetry
with respect to $u\to-u$. The difference between the odd-symmetric
and non-symmetric cases (even symmetry is not allowed
due to the presence of the constant $c$) appears from the way nonlinearity
enters the equation through $P(u)$ and $A(u)$. It is also interesting to
observe that the velocity of the solitons $c$ only depends on the even
coefficients $p_{0}$ and $p_{2}$, and odd coefficients $a_{1}$ and 
$a_{3}$. This implies that, contrarily to the solutions 
discussed in Sec.~II, both diffusion and dispersion effects are actually
relevant to determine $c$:
\be
c=p_{0}-\frac{1}{\lambda a^{2}} ( a_{1}-\nu \lambda^{2} a^{4} 
- \delta \lambda^{3} a^{6} )~.
\ee
We also recall that, as for the previous case, the coefficients are not 
totally independent and relations given by Eqs.~(21) and (22) still
hold for soliton solutions in the form of Eq.~(36). 

Finally, as an illustrative example we consider the case in which
\be
P(u) = p_{0}~,
\ee
and
\be
A(u)=-9\lambda^{3}a^{4}\delta u^{3}+24\lambda^{3}a^{2}\delta u^{5}-
15\lambda^{3}\delta u^{7}~.
\ee
If we set
\be 
\nu = -\lambda a^{2} \delta~,
\ee
then Eq.~(8) with solutions given by Eq.~(36) is equivalent to a gKdVBH
equation, Eq.~(1), defined by the functions
\ben
f(u)& =& f_{0} + p_{0} u~,\\
g(u)& = & g_{0} + \nu u~,\\
h(u)&= & A(u)~,
\een
and with $\bar{\delta} = -\delta$, if the velocity of the chiral 
kink soliton solution, Eq.~(36), is 
\be
c = p_{0}~,
\ee 
which is 
also in agreement with Eq.~(48) and the results reported in Ref.~\cite{ern}.

\section{Conclusion}

In this work we have investigated the presence of travelling
kink soliton solutions $u(x,t) = u(x - ct)$
to nonlinear partial differential equations of the generic form
$$
u_{t} + P(u) u_{x} + \nu u_{xx} +
\delta u_{xxx} = A(u)~,
$$
with polynomial functions defined as
$$
P(u) = \sum_{i=0}^{N_{p}} p_{i}\,u^{i}~,\hspace{1cm}
A(u) = \sum_{i=0}^{N_{a}} a_{i}\,u^{i}~.
$$
These equations combine diffusion, dispersion,
and nonlinearity in distinct ways, and include a number of examples of
relevant differential equations in physics and other fields of nonlinear
science. 

We have analyzed two types of kink soliton solutions,
namely $u(y) = a \tanh(\lambda a y)$, which is also related to solutions
of the relativistic $1+1$ dimensional $\phi^{4}$ field-theoretical system,
and $u(y)=\{(a^{2}/2)[1+\tanh (\lambda a^{2} y)]\}^{1/2}$, which
is associated with kinks of the $\phi^{6}$ system. These kinks present distinct
properties. 
For instance, while in the $\phi^4$ system they connect the two asymmetric
vacua, in the $\phi^6$ model they relate the symmetric vacuum $\phi=0$ to
the asymmetric ones. 

The emphasis of our study was on the chirality of the localized soliton 
solutions, and its relation with diffusion, dispersion, and nonlinear
effects, as well as its dependence on the parity of the polynomials
$P(u)$ and $A(u)$ with respect to the discrete symmetry $u\to-u$.

\section*{Acknowledgments}

EPR and DB would respectively 
like to thank the Condensed
Matter Theory group at Harvard University and 
the Center for Theoretical Physics at 
Massachusetts Institute of Technology for hospitality.

\end{document}